\documentclass[12pt]{iopart}

\usepackage{iopams}

\expandafter\let\csname equation*\endcsname\relax
\expandafter\let\csname endequation*\endcsname\relax

\usepackage{amsmath}
\usepackage{graphicx}
\usepackage{hyperref}
\usepackage{float}
\usepackage[normalem]{ulem}
\usepackage{subcaption}
\usepackage{epstopdf}
\usepackage{color}

\newcommand{\bra}[1]{\ensuremath{\left\langle#1\right|}}
\newcommand{\ket}[1]{\ensuremath{\left|#1\right\rangle}}

\begin{document}

\title[Revealing memory effects in phase-covariant quantum master equations]{Revealing memory effects in phase-covariant quantum master equations}
\author{J Teittinen$^1$, H Lyyra$^1$, B Sokolov$^1$ and S Maniscalco$^{1,2}$}
\address{$^1$ QTF Centre of Excellence, Turku Centre for Quantum Physics, Department of Physics and Astronomy, University of Turku, FI-20014 Turun Yliopisto, Finland}
\address{$^2$ QTF Centre of Excellence, Department of Applied Physics, School of Science, Aalto University, FI-00076 Aalto, Finland}
\ead{jostei@utu.fi}

\begin{abstract}We study and compare the sensitivity of multiple non-Markovianity indicators for a qubit subjected to general phase-covariant noise. For each of the indicators, we derive analytical conditions to detect the dynamics as non-Markovian. We present these conditions as relations between the time-dependent decay rates for the general open system dynamics and its commutative and unital subclasses. These relations tell directly if the dynamics is non-Markovian w.r.t.~each indicator, without the need to explicitly derive and specify the analytic form of the time-dependent coefficients. Moreover, with a shift in perspective, we show that if one assumes only the general form of the master equation, measuring the non-Markovianity indicators gives us directly non-trivial information on the relations between the unknown decay rates. 
\end{abstract}

\maketitle

\section{Introduction}

Open quantum systems theory allows us to investigate the effects of environmental noise on the temporal behaviour of quantum systems. This general theoretical framework permits not only to tackle fundamental open questions, such as the quantum measurement problem, but it also has practical applications to quantum technologies. Modelling and understanding the loss of quantum properties caused by the interaction between quantum devices and their surroundings is, indeed, a pre-requisite to develop strategies to protect them, hence prolonging their quantum-enhanced efficiency. 

The dynamics of open quantum systems can be divided into two main categories based on the specific way in which they exchange information with the environment. Markovian open quantum systems are characterised by a continuous and monotonic loss of information caused by the monitoring action of the environment. Generally, Markovian dynamics occurs when the interaction between system and environment is sufficiently weak and the environment correlations are short-living \cite{breuer_petruccione}. Non-Markovian open quantum systems, on the contrary, display memory effects which manifest themselves as information backflow and/or partial return of previously lost quantum properties \cite{rivas2014, breuer2016,devega2017}. Several different quantifiers of information have been used to describe non-Markovianity in terms of information flow. For a comprehensive review of the different physical interpretations of non- Markovianity and how they relate to each other, see \cite{li2018}.

The theoretical analysis of non-Markovian dynamics is undoubtedly more demanding than the Markovian case, not only because the corresponding master equation is more complicated to solve both analytically and numerically, but also and most importantly because we lack a general theorem guaranteeing the physicality of the mathematical solution of the master equation. In other words a generalization of the well-known Gorini-Kossakowski-Sudarshan-Lindblad (GKSL) theorem \cite{gorini1976, lindblad1976} to non-Markovian systems is still unknown. Nonetheless, with the advance of experimental solid-state platforms for quantum technologies, where the noise is generally non-Markovian, and with the development of sophisticated reservoir engineering techniques, investigations on non-Markovian open quantum systems have become highly topical \cite{liu2011,tang2012,xiang2014,bernardes2015,bernardes2016}.

The modern approach to non-Markovian dynamics focuses on the physical characterization of memory effects and describes them in terms of a partial, and generally temporary, recovery of previously lost information. Within quantum information theory, several quantities are used to measure the amount of information stored in a quantum system, but only some of them satisfy the physical requirements needed to properly describe information flow and memory effects \cite{wilde2013,nielsen2000}. Perhaps the most important of such requirements is the so-called contractivity property according to which, in presence of environmental noise, the amount of information present in the system at the initial time is always greater or equal to the amount of information at a subsequent time $t$. 

Examples of quantifiers that are contractive under completely positive and trace preserving (CPTP) maps are trace distance, relative entropy, Fisher information, quantum mutual information, fidelity, coherent information and so on and so forth \cite{wilde2013,vedral2002,lu2010,rivas2010,bylicka2014}. Consequently, these quantities have been used to define different indicators of non-Markovianity \cite{lu2010,rivas2010,bylicka2014,luo2012,vasile2011}. Whenever the dynamics of such information quantifiers shows non-monotonic behaviour, one interprets the partial return of information as a manifestation of memory effects and says that the time evolution is non-Markovian. These indicators of non-Markovianity, however, do not always coincide and, moreover, they are generally difficult to calculate since they require at least the knowledge of the full solution of the master equation.

In this paper we show that, for the most general form of phase-covariant single qubit dynamics, one can bypass the problem of solving the master equation and directly identify the Markovian or non-Markovian character of the dynamics, as detected by a given indicator, by verifying certain simple inequalities on the decay rates appearing in the master equation. We apply this finding to several recently introduced non-Markovianity indicators and we use this approach to make a comprehensive comparison of their sensitivity. Since our approach does not depend on the specification of the analytic form of the time-dependent decay rates, it can be applied in a variety of physical situations, such as those considered in \cite{chruscinski2017, cosco2017, cianciaruso2017, addis2016, bylicka2016, addis2014, haikka2012}.

The paper is structured as follows. In section \ref{sec:2_mastereq} we introduce the open quantum system model and recall some of its properties. In section \ref{sec:3_nm_indicators} we introduce the non-Markovianity indicators presented and derive some mathematical properties that will be useful to characterise memory effects. Section \ref{sec:4_results} contains our results, namely the inequalities on the decay rates identifying non-Markovianity. Finally in section \ref{sec:5_conclusions} we summarise our results, compare the findings of the previous section and present conclusions.

\section{Master equations and physical models}\label{sec:2_mastereq}

Let us begin by recalling some general definitions and properties on open quantum systems theory. 
We consider a time-local master equation of the form.
\begin{align}\label{me}
\dfrac{\rmd \rho(t)}{\rmd t} = -\frac{i}{\hbar}[\rho(t),H(t)] + \sum_i \gamma_i(t)\bigg(A_i\rho(t) A_i^\dagger -\frac{1}{2}\Big{\{}A_i^\dagger A_i,\rho(t) \Big{\}}\bigg)\,,
\end{align}
where $H(t)$ is the system Hamiltonian, $A_i$ are the Lindblad or jump operators and the time-dependent real-valued functions $\gamma_i(t)$ are the decay rates. The solution of the master equation defines the dynamical map $\rho(t) = \Phi_t ( \rho (0))$, with $\Phi_0 = \mathbb{I}$.

Generally, a map is called $k$-positive if the composite map $\Phi_t \otimes \mathbb{I}_k$, where $k$ is the dimensionality of the ancillary Hilbert space and $\mathbb{I}_k$ its identity operator, is positive. If $\Phi_t \otimes \mathbb{I}_k$ is positive for all $k \geq 0$ and for all $t$, then we say that the dynamical map is completely positive. A dynamical map $\Phi_t$ is called CP-divisible (P-divisible) if the propagator $V_{t,s}$, defined by
\begin{equation}
\Phi_t = V_{t,s}\circ\Phi_s\,,
\end{equation}
is completely positive (positive) for all $t \geq s \geq 0$ \cite{breuer_petruccione}.

Using the GKSL theorem we can immediately draw general conclusions on open quantum systems described by a master equation of the form of equation \eqref{me}: If $\gamma_i(t)$ are constant and positive, the resulting dynamical map is a semigroup and the solution is always physical, i.e., CPTP \cite{gorini1976}. This result still holds for non-negative time-dependent $\gamma_i(t)$, in this case the dynamical map is not a semigroup but CP-divisible \cite{rivas2012}. We can therefore generally infer whether the dynamics is Markovian (CP-divisible) or not directly by looking at the decay rates appearing in the master equation. If at least one of the the decay rates $\gamma_i(t)$ attains, even if temporarily, negative values, then the dynamical map is not CP-divisible and the physicality conditions become more complicated \cite{lankinen2017}.

CP-divisibility has been proposed as the definition for Markovianity in open quantum systems \cite{rivas2014,rivas2010,bylicka2014}. Despite its clear mathematical formulation, in general, detecting the violation of the CP-divisibility condition is not experimentally straightforward. A hierarchy of non-Markovianity indicators, based on the violation of k-divisibility, have been introduced and the extreme case of essentially non-Markovian maps have been experimentally studied \cite{chruscinski2014}. 

In the following we review some non-Markovianity indicators based on quantifiers of information on the system. We recall that, when any of these indicators display non-monotonic behaviour, indicating the presence of memory effects, we can conclude that the dynamical map is not CP-divisible. However, there can be cases in which the indicators display a monotonic behaviour but still the map is not CP-divisible. 

\subsection{Dissipation, heating, and pure dephasing dynamics}

We consider the most general class of master equations describing phase covariant noise. As shown in \cite{smirne2016}, all physical master equations for a single qubit phase covariant dynamics are of the form: 
\begin{equation}\label{lanme}
\dfrac{\rmd \rho(t)}{\rmd t} = -i \dfrac{\omega(t)}{2} [\sigma_z,\rho(t)] + \dfrac{\gamma_1(t)}{2}L_1(\rho(t)) + \dfrac{\gamma_2(t)}{2}L_2(\rho(t)) + \dfrac{\gamma_3(t)}{2}L_3(\rho(t))\,,
\end{equation}
where
\begin{align}
L_1(\rho(t)) &= \sigma_+ \rho(t) \sigma_- - \frac{1}{2} \left\{ \sigma_-\sigma_+, \rho(t) \right\}\,, \\
L_2(\rho(t)) &= \sigma_- \rho(t) \sigma_+ - \frac{1}{2} \left\{ \sigma_+\sigma_-, \rho(t) \right\}\,, \\
L_3(\rho(t)) &= \sigma_z \rho(t) \sigma_z - \rho(t)\,,
\end{align}
where $\sigma_{\pm}= (\sigma_x \pm i \sigma_y)/2$ and $\sigma_x,\,\sigma_y$ and $\sigma_z$ are the Pauli operators. Here $\gamma_1(t)$, $\gamma_2(t)$, and $\gamma_3(t)$ correspond to heating, dissipation and pure dephasing, respectively. In \cite{haase2017}, it was shown, that the general spin-boson time-convolutionless master equation reduces exactly to \eqref{lanme} when applying the secular approximation, without Born-Markov approximation. The necessary and sufficient conditions for complete positivity of the general solutions of equation \eqref{lanme} were studied in \cite{lankinen2017}. For arbitrary single qubit density operator $\rho(0)$, the solution of equation \eqref{lanme} is given by
\begin{equation}\label{lansolu}
\Phi_t(\rho(0)) = \rho(t) = 
\begin{pmatrix}
1 - P_1(t)				& \alpha(t) \\ 
\alpha^*(t)				& P_1(t) 
\end{pmatrix}\,,
\end{equation}
where
\begin{align}
P_1(t) 					&= \rme^{-\Gamma(t)}[G(t) + P_1(0)]\,, \label{eq:P1dyn}\\
\alpha (t)				&= \alpha(0) \rme^{i\Omega(t) - \Gamma(t)/2 - \tilde{\Gamma}(t)}\,,
\end{align}
and\footnote{This solution is trace preserving only for trace one operators. To generalize this to all Hermitian operators, the solution is $P_0(t)=\Tr[\rho(0)]-P_1(t)$ and $G(t)=\Tr[\rho(0)]G(t)$. In this paper, it is sufficient to focus just on trace one preserving maps.}
\begin{align}
\Gamma(t) 				&= \dfrac{1}{2} \int_0^t (\gamma_1(\tau) + \gamma_2(\tau)) d\tau\,,\label{ga} \\
\tilde{\Gamma}(t) 		&= \int_0^t \gamma_3(\tau) d\tau\,,\\
G(t) 					&= \frac{1}{2} \int_0^t \rme^{\Gamma(\tau)} \gamma_2(\tau) d\tau\,,\label{g}\\
\Omega(t)				&= \int_0^t 2 \omega(\tau)d\tau\,.
\end{align}

The master equation of equation \eqref{lanme} leads to commutative dynamics, meaning $\Phi_t \circ \Phi_s = \Phi_s \circ \Phi_t$, for any $s,t \geq 0$, iff $\gamma_1(t) = \gamma(t)$ and $\gamma_2(t) = \kappa \gamma(t)$, where $0\leq \kappa \leq 1$\footnote{In general $\kappa$ could be any positive number. However, we can restrict $\kappa \in [0,1]$, because the case of $\kappa>1$ would again correspond to the choice of $\gamma_1(t) = \kappa \times \gamma(t)$ and $\gamma_2(t) = \gamma(t)$, where $\kappa \in [0,1]$. This choice affects the resulting dynamics, but does not change the results in this paper. Negative $\kappa$ violates the CP conditions of \cite{lankinen2017}}. In the commutative case, equations \eqref{ga} and \eqref{g} simplify to
\begin{align}
\Gamma_\kappa(t) 		&= \dfrac{1+\kappa}{2} \int_0^t \gamma(\tau) d\tau\,,\label{ga_comm} \\
G(t) 					&= \frac{\kappa}{\kappa +1} \Big(\rme^{\Gamma_\kappa(t)} - 1 \Big)\,,\label{g_comm}
\end{align}

We notice the dynamics is unital, meaning $\Phi_t$ satisfying $\Phi_t(1/2\mathbb{I}) = 1/2\mathbb{\mathbb{I}}$, where $\mathbb{I}$ is the identity operator, iff it is commutative and $\kappa=1$. For these choices, the master equation reduces to
\begin{equation}\label{lanmeu}
\dfrac{\rmd \rho(t)}{\rmd t} = -i \dfrac{\omega(t)}{2} [\sigma_z,\rho(t)] + \frac{\gamma(t)}{2}L_x(\rho(t)) + \frac{\gamma(t)}{2}L_y(\rho(t)) + \dfrac{\gamma_3(t)}{2}L_3(\rho(t)),
\end{equation} 
with
\begin{align}
L_x(\rho(t)) 			&= \sigma_x \rho(t) \sigma_x - \rho(t)\,, \\
L_y(\rho(t)) 			&= \sigma_y \rho(t) \sigma_y - \rho(t) \,.
\end{align}
The Lindblad operators in equation \eqref{lanmeu} are the Pauli matrices, and thus the dynamical map is also Hermitian, meaning $\Phi_t^* = \Phi_t$, for all $t>0$, where $\Phi_t^*$ denotes the dual map of $\Phi$. We note that the dynamical map \eqref{lansolu} is Hermitian exactly in the unital case.

\subsection{Always physical phenomenological model}\label{sec:f_model}

For the sake of concreteness, we will consider in the following also a specific model, firstly introduced in \cite{lankinen2017}, which has the nice features of satisfying always the complete positivity conditions and, in certain limits, reproducing the correct Markovian master equation (with positive and constant decay rates). The model assumes a certain time dependency for the three decay rates appearing in 
the master equation~\eqref{lanme}, parametrized by two positive numbers $R$ and $s$. Specifically, the first two decay rates are given by
\begin{align}
\gamma_1(t) 		&= 2N f(t)\,, \\
\gamma_2(t) 		&= 2(N+1)f(t)\,,
\end{align}
where	
\begin{align}
f(t) 		&= -2 \text{ Re} \Big\{ \dfrac{\dot{c}(t)}{c(t)} \Big\}\,, \label{eq:f_gamma1} \\
c(t) 		&= c(0) \rme^{-t/2}\Bigg[ \cosh (\sqrt{1-2R}t/2) + \dfrac{\sinh(\sqrt{1-2R}t/2)}{\sqrt{1-2R}} \Bigg]\,, \label{eq:f_gamma2}
\end{align}
where $R$ depends on both the coupling with the environment and the environmental spectral properties, and $N(T) = [\exp(\hbar \omega_0/kT) -1]^{-1}$ is the mean number of excitations in the modes of the thermal environment, with $T$ the temperature and $\omega_0$ the Bohr frequency .

Note that, for $R<1/2$, the decay rates $\gamma_1(t)$ and $\gamma_2(t)$ are always positive, while for $R>1/2$ they become negative for certain time intervals.
The pure dephasing rate is given by~\cite{lankinen2017}
\begin{equation}\label{eq:f_gamma3}
\gamma_3(t) = 2 \int d\omega J(\omega) \coth (\omega/k_BT) \sin (\omega t)\,,
\end{equation}
where the spectral density $J(\omega)$ is 
\begin{equation}
J(\omega) = \dfrac{\nu \omega^s}{\omega_c^s} \rme^{-\omega/\omega_c}\,,
\end{equation}
with $\omega_c$ the cut-off frequency and $\nu$ a dimensionless coupling constant. For pure dephasing, that is when $\gamma_1(t) = \gamma_2(t) = 0$, markovianity of this model depends on a critical value $s_{crit}(T)$ of the Ohmic parameter $s$, the system being non-Markovian when $s > s_{crit}(T)$. The critical value depends on temperature and increases monotonically between minimum of $s_{crit}(0)=2$ and maximum $s_{crit}(T\rightarrow \infty)=3$ \cite{haikka2013}.

\section{Non-Markovianity indicators}\label{sec:3_nm_indicators}
In this section we give a brief review of the non-Markovianity indicators that we will use in the paper and present some properties which will be needed to derive the main results, presented in section \ref{sec:4_results}.

\subsection{Entropy production rate}\label{sec:changeinentropy}

In non-equilibrium thermodynamics the entropy production is defined as
\begin{equation}
\sigma = \frac{\partial S}{\partial t} + div \mathcal{J},
\end{equation}
where $S$ stands for the entropy of the open system and $\mathcal{J}$ is a vector field describing the flow of entropy per unit area per unit time. In quantum mechanics $S$ is the von Neumann entropy, defined as $S(\rho(t)) = - \tr[\rho(t) \log \rho(t)]$. Entropy production was introduced in the framework of open quantum systems in \cite{spohn1978}, where it was shown that it can be written as 
\begin{equation} \label{eq:entropyflow}
\sigma_\beta(\rho(t)) = - \dfrac{\rmd}{\rmd t} S (\rho(t) \vert \rho_\beta)\Bigr|_{\substack{t=0}}\,,
\end{equation}
where $S(\rho(t) \vert \xi(t)) = -\tr [\rho(t) \log \xi(t)] - S(\rho(t))$ is the quantum relative entropy, $\rho(t)$ is the system state at time $t$ and $\rho_\beta$ is the thermal equilibrium state at temperature $\beta^{-1}$. It was also shown that this could be generalized to any dynamical map $\Phi_t$ with a stationary state $\rho_0$, such that $\Phi_t(\rho_0) = \rho_0, \forall t$. In \cite{Muller2017}, the relative entropy was proven to be contractive for all positive maps, that is $S(\rho_1\vert \rho_2) \leq S(\Phi(\rho_1)\vert \Phi(\rho_2))$, for all positive maps $\Phi$. The contractivity of quantum relative entropy is one the most fundamental inequalities in quantum information theory, and can be intrepreted as a decrease in distinguishability between two states \cite{wilde2013}. Later, the contractivity was used as a basis for non-Markovianity measures \cite{bhattacharya2017,he2017}. For any stationary state $\rho_0$, the quantum entropy production at time $t$ is
\begin{equation}\label{eq:ourentropyflow}
\sigma(\rho(t)) = \dfrac{\rmd}{\rmd t} \tr[\rho(t) \log \rho_0] + \dfrac{\rmd S(\rho(t))}{\rmd t}\,.
\end{equation}

In order to understand the physical meaning of this non-Markovianity indicator we stress again the fact that the quantum relative entropy is a measure of distinguishability between quantum states, even if it is neither symmetric nor does it satisfy triangular inequality. Therefore it does not define a proper metric. Despite these drawbacks it has been extensively used in quantum information theory being the direct generalization of the classical relative entropy (which is also not symmetric and therefore not a metric) \cite{note}. Moreover,  quantum relative entropy is useful since several other important quantities, e.g., quantum mutual information and quantum conditional entropy, are special cases of it and, as such, it can be used to quantify quantum information and entanglement \cite{vedral2016, vedral1997}.

The physical description of memory effects in terms of partial and temporary increase of quantum relative entropy during the time evolution stems from the quantification of  information backflow as partial increase of state distinguishability.  Specifically, entropy production is defined in terms of quantum relative entropy between the system state at time $t$ and the stationary state.  Loss of information, or more precisely information flow, is here quantified as a decrease in the distance to the asymptotic state of the dynamics. Hence memory effects indicate a partial increase in the distinguishability between the state of the system and its stationary state or, equivalently, as a partial increase of information.

For the time evolution of a generic one-qubit state
\begin{equation}
\rho(t) = 
\begin{pmatrix}
1 - \rho_{00}(t)	& \rho_{01}(t) \\ 
\rho_{10}(t)	& \rho_{11}(t) 
\end{pmatrix}\,,
\end{equation}
the eigenvalues of $\rho(t)$ are $\lambda_\pm (t) = (1 \pm x(t))/2$, with $x(t) = \sqrt{4| \rho_{01}(t) |^2 + (2\rho_{11}(t) - 1)^2}$. By the positivity and $\tr [\rho]$ = 1 properties 
of quantum states, we get $0\le x(t)\le 1\,\forall t$. We notice that, for unital dynamics, the first term of equation \eqref{eq:ourentropyflow} becomes zero and the total entropy flow reduces to von Neumann entropy. This means that von Neumann entropy can be used as a non-Markovianity indicator for unital maps, as seen in \cite{bhattacharya2017}. The time derivative of the von Neumann entropy becomes
\begin{equation}
\frac{\rmd S}{\rmd t} = -\frac{\rmd}{\rmd t}\text{tr}[\rho(t) \log \rho(t)] = \dfrac{1}{2} \log \left[ \dfrac{1-x(t)}{1+x(t)} \right] \dfrac{\rmd x(t)}{\rmd t}\,. \label{vonneumann}
\end{equation}
On the other hand, since $0\le x(t)\le 1\,\forall t$, one sees that
\begin{align}
&\log \left[ \dfrac{1-x(t)}{1+x(t)} \right] < 0 \,,
\end{align}
and thus we conclude, that 
\begin{align}\label{voneg}
\begin{aligned}
\dfrac{\rmd S}{\rmd t}<0 
~~\Leftrightarrow~~ \dfrac{\rmd x(t)}{\rmd t}>0  
~~\Leftrightarrow~~ \dot{\rho}_{11}(t)[2\rho_{11}(t) - 1] + \frac{\rmd}{\rmd t}\vert \rho_{01}(t)\vert^2 >0\,.
\end{aligned}
\end{align}

\subsection{Purity rate}

In \cite{bhattacharya2017} the rate of change of purity was considered as a measure of non-Markovianity. Purity, defined as $\mathcal{P}(\rho(t)) = \Tr[\rho(t)^2]$, is generally used as a quantifier of quantumness of a state, and is directly related to other useful quantities such as concurrence \cite{rungta2001}.
If the Lindblad operators $A_i$ in equation \eqref{me} are Hermitian then the purity rate can be written as \cite{bhattacharya2017}
\begin{align}
\dot{\mathcal{P}}(\rho(t))&=
-\sum_i \gamma_i (t) Q_i (t)\,,
\end{align}
where $Q_i(t)=||[A_i,\rho(t)]||^2_{\text{HS}}$ and HS stands for the Hilbert-Schmidt norm, defined for an operator $A$ as $||A||_{\text{HS}}=\sqrt{\tr[A^\dagger A]}$. As $Q_i (t)\ge 0 , \forall t\ge 0$, the positivity of $\dot{\mathcal{P}}(\rho(t))$ guarantees negativity of at least one $\gamma_i (t )$. Generally, the positivity of purity rate cannot be used as non-Markovianity indicator for non-unital channels, since the purity reaches its minimum when the system is in the maximally mixed state.

For generic qubit dynamics, the purity change rate can be written as
\begin{align}
\dot{\mathcal{P}}(t) 
&= 2\bigg{\{} \dot{\rho}_{11}(t)[2\rho_{11}(t) - 1] + \frac{\rmd}{\rmd t}\vert \rho_{01}(t)\vert^2 \bigg{\}}\,.
\end{align}
By looking at equation \eqref{voneg}, we notice that $\dot{\mathcal{P}}(t)>0\Leftrightarrow \rmd S/ \rmd t<0$. In fact, all unital channels have the maximally mixed state as a stationary state, so we see that purity rate indicator is just a more restrictive case of entropy production indicator, since it can be properly used as non-Markovianity witness only for unital maps. Therefore, for this class of dynamical maps, memory effects can also be interpreted as a partial return of quantumness as indicated by purity of the open quantum system state.

\subsection{Trace distance}

A commonly used distance measure for quantum states, namely trace distance, is defined for two states, $\rho_1$ and $\rho_2$, as
\begin{equation}
D(\rho_1,\rho_2) = \frac{1}{2} \tr |\rho_1 - \rho_2|\,,
\end{equation}
where $|\cdot|$ indicates absolute value, defined as $|A|=\sqrt{A^\dagger A}$, for some operator $A$. $D(\rho_1,\rho_2)$ is contractive under positive and trace-preserving maps, and thus an increase of $D(\rho_1,\rho_2)$ implies violation of P-divisibility (and, hence, CP-divisibility). Moreover, trace distance has an important role in quantum information: if Alice prepares a system in either state $\rho_1$ or $\rho_2$ with equal probability and sends it to Bob who has to discriminate between the two, the maximum probability of correctly identifying the received state with a single measurement is $(1+D(\rho_1,\rho_2))/2$ \cite{gilchrist2005}. Since state distinguishability can be interpreted as a measure of the amount of information we have on a quantum system, this property, together with contractivity, was used to define the increase of trace distance as a signature of non-Markovianity (BLP measure) and to physically identify the corresponding memory effects as information backflow \cite{breuer2009}. We note that this physical interpretation is similar to the one presented in section 3.1 for the entropy production rate, since both these quantities are based on the description of information flow in terms of dynamical change of state distinguishability (quantified by either trace distance or quantum relative entropy).

The dynamics of a qubit state can be written as the equation of motion for the corresponding Bloch vector $\mathbf{r}(t)$, as $\dot{\mathbf{r}}(t) = \mathcal{D}(t) \mathbf{r}(t) + \mathbf{v}(t)$, where $\mathcal{D}(t)$ is called the damping matrix of the map, and $\textbf{v}(t)$ the drift vector. In our case $\textbf{v}(t) = 0$. It was shown in \cite{Hall2014}, that the trace distance between two states can increase iff
\begin{equation}
\lambda_{max}[\mathcal{D}(t)^T+\mathcal{D}(t)] > 0\,.\label{eq:tracecondition}
\end{equation}
where $\lambda_{max}[A]$ means the maximum eigenvalue of some operator $A$ and $\mathcal{D}(t)^T$ is the transpose of $\mathcal{D}(t)$.

\subsection{Bloch volume measure}

Using the Bloch vector representation of qubit states, one can also investigate the dynamics of an open quantum system by looking at the volume of physical states that are dynamically accessible to the system. We refer to this quantity as the Bloch volume. The Bloch volume is contractive under positive and trace preserving maps and therefore can be used as a geometric indicator of non-Markovianity~\cite{lorenzo2013}. More precisely, an increase in the Bloch volume signals violation of P-divisibility. 
In \cite{Hall2014}, it was shown, that the time evolution of the Bloch volume depends directly on the trace of the damping matrix $\mathcal{D}(t)$. The Bloch volume increases iff
\begin{equation}\label{eq:blochcondition}
\text{tr}[\mathcal{D}(t)]>0\,,
\end{equation}
and thus, if there exist an interval of time such that equation (\ref{eq:blochcondition}) is satisfied, the dynamics is non-Markovian w.r.t.~the Bloch volume measure. By comparing this result with the condition of equation \eqref{eq:tracecondition}, we see that the increase in trace distance is always a stronger indicator for non-Markovianity than the increase of Bloch volume \cite{Hall2014}.

The change in Bloch volume is directly linked to the BLP measure. In the case of a qubit, the optimal pair of states used in the definition of the BLP measure resides at the boundary of the state space. Trace distance in this case is just the Euclidean distance between the two states. This leads to an increase (decrease) in trace distance between these two pairs when the Bloch volume increases (decreases). Despite the similarities, however, this connection can not be used to interpret the change in Bloch volume as change in distinguishability, as discussed in \cite{lorenzo2013}.
The Bloch volume indicator, however, can be interpreted as a measure of the change of classical information encoded in the quantum state. Given a quantum state with probability distribution $p(\textbf{r}_t)$, where $\textbf{r}_t$ is the Bloch vector characterizing the state at time $t$, the change in Shannon entropy $h$ can be written as $h(p(\textbf{r}_t)) - h(p(\textbf{r}_0)) = \log_2 || \textbf{A}_t ||$, where $|| \textbf{A}_t ||$ is the norm of the Bloch vector dynamical map defined as $\textbf{r}_t = A_t \textbf{r}_0 + \textbf{q}_t$. The latter quantity is directly linked to the change in Bloch volume. Thus the decrease in Bloch volume implies a decrease in the amount information  as detailed in \cite{lorenzo2013}. 
According to this interpretation, memory effects are temporary revival of classical information encoded in the quantum state, as it evolves due to the interaction with the environment.

\subsection{Non-Markovianity and singular values of the dynamical map}

Using Bloch volume as an indicator of non-Markovianity was further developed in \cite{chruscinski2017}. The connection between P-divisibility and the changes of Bloch volume was studied using the singular value decomposition. The matrix representation of a map can be written in its singular value decomposition, which consists of two rotation matrices and a single scaling matrix. The scaling matrix is a diagonal matrix consisting of the singular values of the map. Because rotations do not change the Bloch volume, the only relevant part of the map is the scaling matrix. The time evolution of singular values and eigenvalues directly determine if the map is P-divisible or not in the unital and commutative cases respectively.

For unital maps, with singular values $s_k(t)$, P-divisibility is violated iff
\begin{equation}
\frac{\rmd}{\rmd t}s_k(t) > 0\,,
\end{equation}
for at least one $k$, which implies non-Markovianity. For commutative maps, the P-divisibility is determined by the eigenvalues of the map. Let $\lambda_k(t)$ be the eigenvalues of the dynamical map $\Phi_t$, meaning $\Phi_t(X_k)=\lambda_k(t)X_k$ w.r.t.~the operator eigenbasis $\{X_k\}$ of the dynamical map. In this case, P-divisibility is violated iff
\begin{equation}
\dfrac{\rmd}{\rmd t}|\lambda_k(t)| > 0\,,
\end{equation}
which implies non-Markovianity.

We note that, in general, the change in the singular values of the dynamical map does not have a simple physical meaning which would allow us to interpret memory effects associated to their partial increase during the time evolution. However, they are stronger witnesses of non-Markovianity when compared to the Bloch volume measure and, for this reason, we have considered them in this study.

\subsection{\texorpdfstring{$l_1$}- norm measure of coherences}

By definition, all coherence measures are non-increasing under incoherent CPTP (ICPTP) maps, i.e. CPTP maps that preserve diagonal states as diagonal \cite{baumgratz2014}. The revival of coherences can be used as an indicator of non-Markovian dynamics \cite{he2017}. Coherences are also linked to the amount of quantum information in a system and thus a positive rate of change in coherences can be linked to information backflow.

A simple measure of coherences, namely the $l_1$-norm, is defined as
\begin{equation}
C_{l_1}(\rho(t)) = \sum_{i \neq j} |\rho_{ij}(t)|\,,
\end{equation}
where $\rho_{ij}(t)$ indicates the $(i,j)$-element of the density matrix $\rho(t)$ \cite{baumgratz2014}. As a measure of coherences, $C_{l_1}$ is non-increasing under ICPTP maps. Using this property, non-Markovianity w.r.t.~$C_{l_1}$ can be detected iff \cite{he2017}
\begin{equation}
\dfrac{\rmd}{\rmd t} C_{l_1}(\rho(t)) >0\,,
\end{equation}
For one qubit systems, this condition is equivalent to
\begin{equation}\label{eq:l1reduce}
\frac{\rmd}{\rmd t}|\rho_{01}(t)| > 0\,.
\end{equation}

\subsection{Relative entropy of coherences}

Another coherence measure, namely the relative entropy of coherences (REC), is defined as the smallest quantum relative entropy between the state $\rho(t) = \Phi_t(\rho(0))$ and an incoherent state $\xi(t) = \Phi_t(\xi(0))$
\begin{equation}
C_\text{r}(\rho(t)) = \min_{\xi(t) \in \mathcal{I}} S(\rho(t)|\xi(t)) = S(\rho_{diag}(t)) - S(\rho(t))\,,
\end{equation}
where $\rho(t)$ is the system state, and $\rho_{diag}(t)$ is the system density matrix $\rho(t)$, where off-diagonal elements are replaced with zeroes \cite{baumgratz2014}. Like entropy production, REC is also defined using the relative entropy, which was shown to be contractive under (incoherent) positive maps \cite{Muller2017}, and thus revival of coherences implies violation of P-divisibility. In \cite{he2017}, this property was used to define non-Markovianity as increase in coherences, indicated by
\begin{equation}\label{eq:reccondition}
\dfrac{\rmd}{\rmd t} C_\text{r}(\rho(t)) > 0\,.
\end{equation}
We note here, that for one qubit systems, choosing the initial state, so that the diagonal elements are invariant in the dynamics, simplifies the equation \eqref{eq:reccondition} to equation \eqref{eq:l1reduce}.

\section{Results}\label{sec:4_results}

In the previous section, we focused on general single qubit systems. In this section, we concentrate on the general phase covariant single qubit dynamics rising from master equations of the form \eqref{lanme} and its commutative and unital special cases.

\subsection{Entropy production rate}

To calculate the entropy flow, we find a stationary state for the system. For any stationary state, $P_1(0)=P_1(t)$ $\forall t \geq 0$. Using equation \eqref{eq:P1dyn}, this is equivalent to
\begin{align}
\rme^{-\Gamma(t)} G(t) &= P_1(0) (1-\rme^{-\Gamma(t)})\,.
\end{align}
We see that this holds iff the dynamics is commutative. Thus the commutative case is the most general subclass of master equation \eqref{lanme}, for which the entropy production is a valid indicator of non-Markovianity. For each $\kappa$, the corresponding stationary state is
\begin{equation}
\rho_\kappa = \left( \begin{array}{cc}
\dfrac{1}{\kappa + 1} &0 \\ 0& \dfrac{\kappa}{\kappa + 1}
\end{array} \right)\,,
\end{equation}
and the entropy production becomes
\begin{equation}\label{eq:commutativeentropy}
\sigma(t) = \dot{P}_1(t) \log(\kappa) + \dfrac{\rmd S(\rho(t))}{\rmd t}\,.
\end{equation}
The magnitude of $\kappa$ dictates how much entropy production deviates from the von Neumann entropy. For any initial state with the same diagonal elements as the stationary state $\rho_\kappa$, $\dot{P}_1(t)=0$, and the entropy production reduces to von Neumann entropy. Additionally, when $\kappa=1$, the dynamics is unital and the first term disappears, and the entropy production reduces to von Neumann entropy, for any state $\rho(t)$. Next we find the optimal states for detecting non-Markovianity w.r.t.~entropy production.

We can emphasize the off-diagonal dynamics by choosing the same diagonal elements as in the stationary state, and non-zero off-diagonal elements $\alpha(0)$, for the initial state. With this choice, $\dot{P}_1(t)=0$, and thus the first term in equation \eqref{eq:commutativeentropy} is zero. From equation \eqref{voneg}, we see that now the entropy production is negative, iff
\begin{align}
\frac{\rmd}{\rmd t} |\alpha(t)|^2 > 0
~~\Leftrightarrow~~ (1+\kappa)\gamma(t) + 4\gamma_3(t) &< 0\,.\label{eq:commentropy1}
\end{align}

On the other hand, we can emphasize the role of diagonal elements by choosing a diagonal state, so that $P_1(0)=1/(1+\kappa)$ and $\alpha(0)=0$. With this initial state, the entropy production becomes
\begin{equation}
\sigma(t) = \frac{ \kappa - 1 }{2} \gamma(t) \rme^{-\Gamma_\kappa(t)} \log\bigg[ \kappa \frac{1 - (1-\kappa) \rme^{-\Gamma_\kappa(t)}}{\kappa + (1-\kappa)\rme^{-\Gamma_\kappa(t)}} \bigg].
\end{equation}
Since, $0 \leq \kappa \leq 1$, we see that the first factor and the logarithmic function is always negative, and thus the dynamics is non-Markovian w.r.t.~entropy production indicator, when $\gamma(t) < 0 \Leftrightarrow \sigma(t) < 0$. In summary, the conditions for detecting non-Markovianity w.r.t.~entropy production are given by
\begin{align}
(1+\kappa)\gamma(t) + 4\gamma_3(t) &< 0 \,, \\
\gamma(t) &< 0 \,.
\end{align}
We note, that these results apply only to the commutative case of master equation \eqref{lanme}, since stationary states do not exist for the non-commutative cases.

\subsection{Purity rate}\label{purspec}

From the general form of a master equation in the Lindblad form, we notice that Hermicity of the Lindblad operators implies unital dynamics. By looking at equation \eqref{lanmeu}, we see that the Lindblad operators are always Hermitian in the unital case of the master equation \eqref{lanme}. So we see, that for the unital case in equation \eqref{lanmeu}, we can use the positivity of $d\mathcal{P}(t)/dt$ as indicator of non-Markovianity.

For an arbitrary initial qubit state, the purity rate becomes
\begin{align}
\dot{\mathcal{P}}(t) 
 =& -4 \rme^{-\Gamma_1(t)}\Bigg{\{}\gamma(t)\rme^{-\Gamma_1(t)}\Big(P_1(0) - \frac{1}{2}\Big)^2 + \bigg( \frac{\gamma(t)}{2} + \gamma_3(t)\bigg)\vert\alpha(0)\vert^2 \rme^{-\tilde{2\Gamma}(t)} \Bigg{\}}\,.\label{purideri}
\end{align}
Here the decay rates $\gamma(t)$ and $\gamma_3(t)$ might take negative values, but all the other functions appearing in the r.h.s.~of equation \eqref{purideri} are always non-negative. Next we find the optimal states for detecting non-Markovianity w.r.t.~purity rate.

As in the case of entropy production, we can appropriately choose an initial state, in order to emphasize either off-diagonal dynamics, or diagonal dynamics. By choosing $\rho(0) = \ket{+}\bra{+},$ where $\ket{+} = 1/ \sqrt{2}\big(\ket{0} + \ket{1}\big)$, the first part vanishes and $d\mathcal{P}(t)/dt>0$ iff
\begin{equation}\label{eq:purityineq1}
\gamma(t) + 2 \gamma_3(t) < 0\,.
\end{equation}
On the other hand by choosing any diagonal initial state, $d\mathcal{P}(t)/dt>0$ iff
\begin{equation}\label{eq:purityineq2}
\gamma(t) < 0\,.
\end{equation}
 The conditions \eqref{eq:purityineq1} and \eqref{eq:purityineq2} define the boundaries for detecting non-Markovianity with purity rate. As expected, these are the same conditions as in the case of entropy production, as these indicators coincide for unital dynamics.

\subsection{Trace distance}

For the general form of master equation \eqref{lanme}, the damping matrix $\mathcal{D}(t)$ takes the form
\begin{equation}\label{eq:generaldampingmatrix}
\mathcal{D}(t) = \left( \begin{array}{ccc}
-\frac{\gamma_1(t)}{4} -\frac{\gamma_2(t)}{4} -\gamma_3(t)  &  -\omega(t) &  0 \\
\omega(t)   &  -\frac{\gamma_1(t)}{4} -\frac{\gamma_2(t)}{4} -\gamma_3(t) &  0 \\
0  &  0  &  -\frac{\gamma_1(t)}{2} -\frac{\gamma_2(t)}{2}
\end{array} \right)\,,
\end{equation}
and thus
\begin{equation}
\begin{aligned}
\mathcal{D}(t)^T + \mathcal{D}(t) = diag[ 
&-(\gamma_1(t) + \gamma_2(t)+4\gamma_3(t))/2\,,\\&
-(\gamma_1(t) + \gamma_2(t)+4\gamma_3(t))/2\,,\,
-(\gamma_1(t)+\gamma_2(t))
]\,.
\end{aligned}
\end{equation}
Now the condition of equation \eqref{eq:tracecondition} becomes
\begin{align}
\gamma_1(t) + \gamma_2(t)+4\gamma_3(t) &< 0\,, \label{eq:traceineq1}\\
\gamma_1(t) + \gamma_2(t) &< 0 \,.\label{eq:traceineq2}
\end{align}
If either of these holds, then the dynamics is detected as non-Markovian by a trace distance based indicator. We note, that the first condition involves all the decay rates. This means that the dynamics is not detected as non-Markovian w.r.t.~the first condition in cases where one or two the decay rates are negative and others positive and large enough to compensate the negativity. On the other hand, the second condition depends on relative magnitudes of the dissipation and absorption rate but does not depend on the pure dephasing rate. This means that the dynamics is not detected as non-Markovian w.r.t.~the second condition if one of the decay rates is positive and large enough to compensate the possible negativity of the other. In the commutative and unital cases, the first condition depends only on the balance of $\gamma(t)$ and $\gamma_3(t)$, while the second condition depends only on the sign of $\gamma(t)$.

\subsection{Bloch volume measure}

For the general master equation \eqref{lanme}, the Bloch volume measure detects dynamics as non-Markovian, iff
\begin{equation}\label{eq:blochvolineq}
\tr [\mathcal{D}(t)] > 0 ~~\Leftrightarrow~~ \gamma_1(t)+\gamma_2(t)+2 \gamma_3(t) < 0\,.
\end{equation}
As expected, this condition is always weaker than the two given by trace distance. If $\gamma_1(t) + \gamma_2(t)<0$, then trace distance immediately detects the non-Markovianity, because of equation \eqref{eq:traceineq2}, but depending on $\gamma_3(t)$, the Bloch volume indicator might not detect it. On the other hand, if the non-Markovianity rises from the negativity of $\gamma_3(t)$, then equation \eqref{eq:traceineq1} detects it sooner than the condition in equation \eqref{eq:blochvolineq}, since the contribution of $\gamma_3(t)$ is greater in trace distance.

\subsection{Non-Markovianity and singular values of the dynamical map}

For master equation \eqref{lanme}, the unital master equations are a subset of the commutative case, so it is sufficient to study only the eigenvalues of the map. In this case, increase in absolute values of the eigenvalues of the dynamical map indicates non-P-divisibility, and thus non-Markovianity. The eigenvalues of the dynamical map $\Phi_t$ are
\begin{align}
\lambda_1(t) = \lambda_2(t)^* = \rme^{i\Omega(t) - \Gamma_\kappa(t)/2 - \tilde{\Gamma}(t)}\,,~~~~~~ \lambda_3(t) = \rme^{-\Gamma_\kappa(t)}\,,~~~~~~ \lambda_4(t) = 1\,.
\end{align}
As $\lambda_4$ is constant, it does not give any relevant conditions. The dynamical map is non-P-divisible iff
\begin{align}
\dfrac{\rmd}{\rmd t}|\lambda_1(t)| = \dfrac{\rmd}{\rmd t}|\lambda_2(t)| >0 &~~\Leftrightarrow~~ (1+\kappa)\gamma(t) + 4\gamma_3(t) < 0\,, \label{eq:commutativeeigen1}\\
\dfrac{\rmd}{\rmd t}|\lambda_3(t)| > 0 &~~\Leftrightarrow~~ \gamma(t) < 0\,. \label{eq:commutativeeigen2}
\end{align}
We conclude, that non-Markovianity w.r.t.~violation of P-divisibility is detected if either of these conditions holds. We notice that these are exactly the same conditions as in the commutative case of the trace distance indicator, namely \eqref{eq:traceineq1} and \eqref{eq:traceineq2}.

\subsection{\texorpdfstring{$l_1$}- norm measure of coherences}

To study the coherence measures, we use the computational basis $\{ \ket{0},\ket{1} \}$ for qubit dynamics. In this basis, the solution \eqref{lansolu}, for the general master equation maps diagonal states to diagonal states and thus the coherence measures can be used. In the general case, the derivative of $l_1$-norm measure becomes
\begin{equation}
\dfrac{\rmd}{\rmd t} C_{l_1}(\rho(t)) = -(\gamma_1(t)/2 + \gamma_2(t)/2 + 2\gamma_3(t))|\alpha(0)| \rme^{-\Gamma(t)/2 - \tilde{\Gamma}(t)}\,.
\end{equation}
The dynamics is non-Markovian w.r.t. the $l_1$-norm of coherences, iff
\begin{equation}\label{eq:l1positivity}
\dfrac{\rmd}{\rmd t} C_{l_1}(\rho(t))>0 ~~\Leftrightarrow~~ \gamma_1(t) + \gamma_2(t) + 4\gamma_3(t) < 0\,.
\end{equation}
This coincides with the first non-Markovianity condition \eqref{eq:traceineq1} of trace distance measure, but does not say anything about the relation of $\gamma_1(t)$ and $\gamma_2(t)$ without $\gamma_3(t)$. However, this is not the same condition as for the Bloch volume measure, since the weight factor of $\gamma_3(t)$ is different. The $l_1$-norm measure is more reactive to the negativity of $\gamma_3(t)$, but the Bloch volume measure is more effective when detecting the negativity of $\gamma_1(t) + \gamma_2(t)$.

\subsection{Relative entropy of coherences}

Finally we study the REC measure. The time derivative of REC for a general qubit state is
\begin{equation}
\dfrac{\rmd}{\rmd t} C_\text{r}(\rho(t)) = \dot{P_1}(t)  \log \bigg[ \frac{1-P_1(t)}{P_1(t)} \bigg] - \dfrac{1}{2} \dfrac{\rmd x(t)}{\rmd t} \log \left[ \dfrac{1-x(t)}{1+x(t)} \right]\,.
\end{equation}
We see, that in general, the positivity of this depends on the initial state and all of the decay rates, which makes it impossible to deduce any non-Markovianity conditions without simplifying the system.

In the commutative case, the stationary state can be used to simplify REC. By choosing the initial state $\rho(0)$, so that the diagonal elements are the same as the stationary state, we have $\dot{P}_1(t)=0$ $\forall t \geq 0$. Thus we are left only with the off-diagonal evolution, which leads to the same positivity condition as in equation \eqref{eq:l1positivity} for $dC_\text{r}(\rho(t))/dt$. Thus we conclude, that for the commutative case, these two indicators are equivalent.

In the general case, we can do numerical analysis to compare REC with the $l_1$-norm measure. According to numerical calculations, using as an example system $\gamma_1(t)=b(\sin(t) +1)$, $\gamma_2(t)=a$ and $\gamma_3(t)=\sin (\tau t)$ (non-unital, non-commutative), with multiple values of $a,b$, and $\tau$, REC is never stronger indicator for non-Markovianity than $C_{l_1}$. Also, according to numerical analysis, the optimizing state for the REC is $\ket{+}\bra{+}$, which is to be expected, as this state has the maximal off-diagonal elements, in the $\{ \ket{0}, \ket{1} \}$ basis, and REC is a measure of coherences. We also found that in this case, the introduction of an ancilla system, as suggested in \cite{he2017}, does not make the indicator any more sensitive to non-Markovianity, but only increases the amplitude of the time derivative of the REC. However, both REC and $C_{l_1}$ are weaker indicators than trace distance in general, as trace distance produces an additional condition from equation \eqref{eq:traceineq2}.

\section{Conclusions and discussions}\label{sec:5_conclusions}

In Table \ref{table:resulttable}, we summarize the conditions for non-Markovian dynamics corresponding to each indicator, for each case of the master equation. These conditions enable us to deduce the non-Markovianity of the master equation directly from its decay rates. This is useful when we want to design a master equation that exhibits non-Markovianity w.r.t. specific measure of non-Markovianity, or we are given a master equation and we need to check if it produces Markovian or non-Markovian dynamics w.r.t.~different indicators. Also, measuring different indicators gives us now relevant information about the decay rates: for example, measuring Markovian dynamics w.r.t.~Bloch volume measure and non-Markovian dynamics w.r.t.~trace distance at time $t$, we know immediately that $\gamma_3(t) \ge -(\gamma_1(t) + \gamma_2(t))/2 > 0$ at that time. Similar reasoning can be used to approximate the unknown value of $\kappa$ in the commutative case.

Only trace distance, Bloch volume and $C_{l_1}$ indicators can be used at the general level, the other indicators need restrictions on the master equation. The REC based indicator also works on a general level, but can only be evaluated numerically and thus does not produce clear analytical conditions for the decay rates.

If we focus on the unital case, where all indicators are valid, we see that entropy production, purity rate, trace distance, and the eigenvalues and singular values of the dynamical map produce exactly the same conditions and at this level can be regarded as the strongest indicators. The overall condition for non-Markovianity w.r.t.~the indicator with two conditions can be written as $\min\{\gamma_1(t) + \gamma_2(t) + 4\gamma_3(t),\,\gamma_1(t) + \gamma_2(t)\} < 0$. Bloch volume and $C_{l_1}$ (and REC) are clearly weaker than other indicators, but are not equal to one another, or even comparable: Bloch volume is the stronger of these when the negativity of the decay rates comes from $\gamma(t)$ and $C_{l_1}$ is stronger when $\gamma_3(t)$ is negative.

Regarding the Bloch vector evolution, from Table~\ref{table:resulttable}, we see that all the indicators with two conditions can identify non-Markovian dynamics from any direction of Bloch vector dynamics. For example, the singular value indicator detects all directions in the Bloch dynamics, while the Bloch volume measure only detects non-Markovianity by looking at the volume, that is all directions simultaneously. Coherence measures intuitively indicate only the non-Markovianity arising from off-diagonal dynamics. 

\begin{table}[H]
\footnotesize
\[
\begin{array}{|c|c|c|c|}\hline
\text{Indicator} &\text{General} &\text{Commutative, } &\text{Unital, } \\
 & &\gamma_1(t)=\gamma(t), \gamma_2(t)=\kappa\gamma(t) &\gamma_1(t) = \gamma_2(t) = \gamma(t) \\ \hline
\text{Entropy production} 	&- &(1+\kappa)\gamma(t) + 4\gamma_3(t) < 0 &\gamma(t) + 2\gamma_3(t) < 0 \\ 
							&- &\gamma(t) < 0 &\gamma(t) < 0 \\ \hline
 							
\text{Purity} 				&- &- &\gamma(t) + 2\gamma_3(t) < 0 \\
			 				&- &- &\gamma(t) < 0 \\ \hline
 							
\text{Trace distance} 		&\gamma_1(t) + \gamma_2(t) + 4 \gamma_3(t) < 0 &(1+\kappa)\gamma(t) + 4\gamma_3(t) < 0 &\gamma(t) + 2\gamma_3(t) < 0 \\
							&\gamma_1(t) + \gamma_2(t) < 0 &\gamma(t) < 0 &\gamma(t) < 0 \\ \hline
 							
\text{Bloch volume} 		&\gamma_1(t) + \gamma_2(t) + 2 \gamma_3(t) < 0 &(1+\kappa)\gamma(t) + 2 \gamma_3(t) < 0 &\gamma(t) + \gamma_3(t) < 0 \\
 							&  &  &  \\ \hline
 							
\text{Eigenvalues} 			&- &(1+\kappa)\gamma(t) + 4\gamma_3(t) < 0 &\gamma(t) + 2\gamma_3(t) < 0 \\
 							&- &\gamma(t) < 0 &\gamma(t) < 0 \\ \hline
 							
\text{Singular values} 		&- &- &\gamma(t) + 2\gamma_3(t) < 0 \\
 							&- &- &\gamma(t) < 0 \\ \hline
 							
C_{l_1}						&\gamma_1(t) + \gamma_2(t) + 4\gamma_3(t) < 0 &(1+\kappa)\gamma(t) + 4\gamma_3(t) < 0 &\gamma(t) + 2\gamma_3(t) < 0 \\
 							& & & \\ \hline
\end{array}
\]
\caption{Conditions for detecting non-Markovianity with different indicators and different classes of master equations. The commutative and unital results for entropy production, as well as purity, require specific choices of initial states. The upper condition involving $\gamma_3(t)$, was obtained with the initial state defined by $P_1(0)=\kappa/(\kappa+1)$ and $\alpha(0) \neq 0$, while the lower result uses initial state defined by $P_1(0)=1/(\kappa + 1)$ and $\alpha(0)=0$.}\label{table:resulttable}
\end{table}

\begin{figure}
    \centering
    \begin{subfigure}[b]{\textwidth}
        \centering\includegraphics[width=0.75\textwidth]{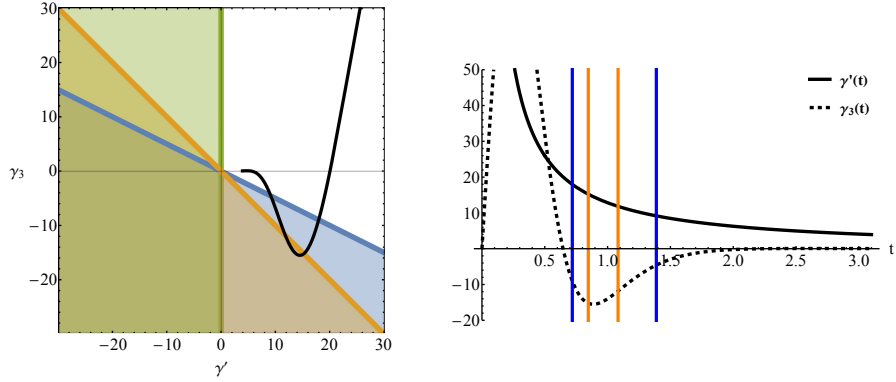}
        \caption{Dynamics of the system in section \ref{sec:f_model}, with R = 0.4, s = 4.5, kT = 3.0}
        \label{fig:fig1a}
    \end{subfigure}\\
    \begin{subfigure}[b]{\textwidth}
        \centering\includegraphics[width=0.75\textwidth]{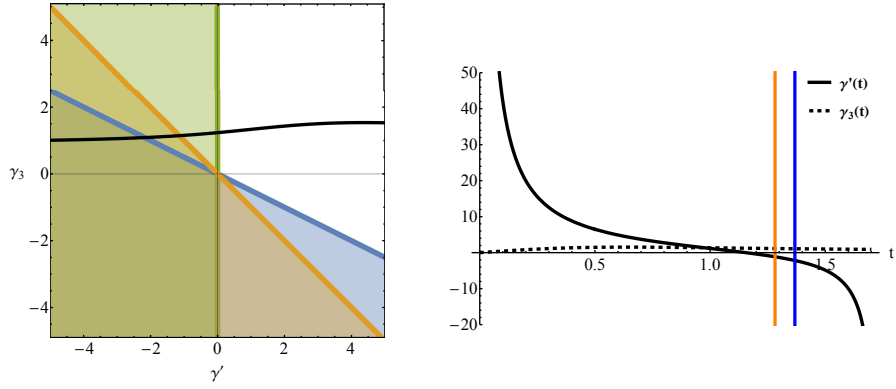}
        \caption{Dynamics of the system in section \ref{sec:f_model}, with R = 4.0, s = 1.0, kT = 3.0}
        \label{fig:fig1b}
    \end{subfigure}\\
    \begin{subfigure}[b]{\textwidth}
        \centering\includegraphics[width=0.75\textwidth]{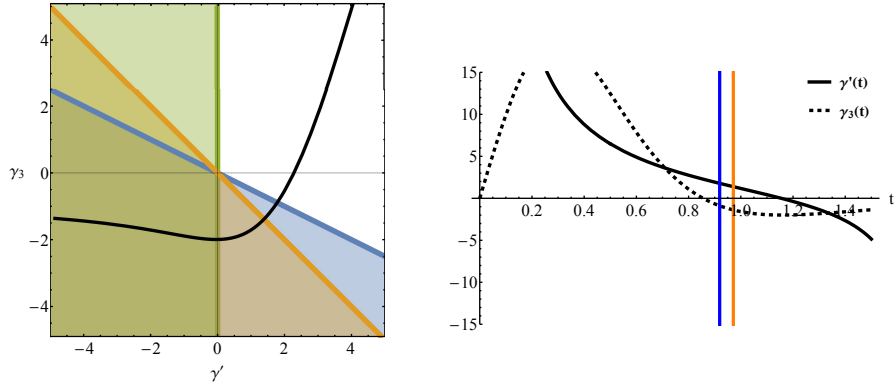}
        \caption{Dynamics of the system in section \ref{sec:f_model}, with R = 4.0, s = 3.5, kT = 3.0}
        \label{fig:fig1c}
    \end{subfigure}
    \caption{\footnotesize On the left-hand side are pictured the regions for each condition from Table \ref{table:resulttable} of the commutative case. The black curve indicates the evolution of the decay rates of the system described in section \ref{sec:f_model} in the $\{ \gamma_3, \gamma' \}$ -space, where we have defined $\gamma'(t) \equiv \gamma_1(t) + \gamma_2(t)$. Here we have chosen $\omega_c = 1$, $\omega_0=1$ and $\nu = 1$. Green region indicates $\gamma'(t) < 0$, blue region $\gamma'(t) + 4 \gamma_3(t) < 0$ and the orange region indicates $\gamma'(t) + 2\gamma_3(t) < 0$. On the right-hand side we have the dynamical plots for $\gamma_3(t)$ (dashed black) and $\gamma'(t)$ (solid black). Vertical lines indicate when the total dynamics crosses border of corresponding color in the left-hand side plots. The crossing to the green region is trivially indicated by the $\gamma'(t)$-curve having negative values.}\label{fig:plottable}
\end{figure}

In Figure \ref{fig:plottable}, the conditions for the commutative case are illustrated as a region plot in $\{ \gamma_3, \gamma' \}$ -space, where $\gamma'(t) \equiv \gamma_1(t) + \gamma_2(t)$. From the regions, we see that all indicators detect non-Markovianity when all decay rates are negative. The transition between Markovian and non-Markovian dynamics happens when the decay rates take negative and positive values at different times. Most of the indicators detect the negativity of $\gamma(t)$, regardless of $\gamma_3(t)$, right away, but the detection of negativity of $\gamma_3(t)$ always depends on the value of $\gamma(t)$. The curves in Figure \ref{fig:plottable} are the dynamics given by the system described in section \ref{sec:f_model}. On the left-hand side are the region plots, with the (black) dynamical curve given by equations \eqref{eq:f_gamma1}, \eqref{eq:f_gamma2} and \eqref{eq:f_gamma3}. On the right hand side we have the time evolution of $\gamma'(t)$ (solid black) and $\gamma_3(t)$ (dashed black) and the vertical lines indicating when the crossing between the regions happen in the left picture.

We also see, that in the lower right region, there exists an area in which none of the indicators detects non-Markovianity. In fact, for any bounded $\gamma_3(t)$, we can choose $\gamma_1(t)$ and $\gamma_2(t)$, s.t. $\gamma_1(t) + \gamma_2(t) + 4\gamma_3(t) \geq 0, \forall t$ and $\gamma_1(t)+\gamma_2(t) > 0$. This means, that violation of CP-divisibility caused by $\gamma_3(t)$ of any magnitude can be hidden from the indicators studied here by positivity of $\gamma_1(t) + \gamma_2(t)$. A special case of these choices is the Eternal non-Markovianity dynamics, introduced in \cite{Hall2014}, which is a special case of master equation \eqref{lanme}, with the choices $\gamma_1(t) = \gamma_2(t) = 2$ and $\gamma_3(t) = -\tanh (t)$. Similarly, the negativity of $\gamma_1(t)$ ($\gamma_2(t)$) can be overshadowed by the positivity of $\gamma_3(t)$ and $\gamma_2(t)$ ($\gamma_1(t)$). Hence the violation of CP-divisibility, rising from negativity of one fixed decay rate, cannot be detected by these indicators if the other rates are chosen accordingly.

In conclusion, in this paper we have thoroughly investigated how to identify non-Markovian dynamics, as signalled by different indicators, by looking at the behaviour of the decay rates of the master equations. Our results apply to a wide class of time-local single-qubit master equations and, as such, are of clear interest for fundamental studies of open quantum systems theory, especially in the context of quantum reservoir engineering.

\section*{Acknowledgements}

The authors acknowledge financial support from the Horizon 2020 EU collaborative project QuProCS (Grant Agreement 641277), the Academy of Finland Centre of Excellence program (Project no. 312058) and the Academy of Finland (Project no. 287750). H.L. acknowledges also the financial support from the University of Turku Graduate School (UTUGS). B.S. acknowledges also the financial support from the Finnish cultural foundation. The authors also acknowledge the helpful discussion with Jyrki Piilo.

\section*{Author contributions}

All authors contributed in planning the research, analyzing and discussing the results and improving the manuscript. J.T. and H.L. wrote the manuscript with the help of S.M.. J.T. and H.L. provided the analytical calculations. B.S. provided the numerical data and plots for Figure 1.

\section*{References}



\begin{thebibliography}{10}
\bibitem{note}
Relative entropy can be given an information theoretic interpretation in the classical case. In source coding, relative entropy can be used to describe inefficiency in coding with random variables with an incorrect information about the original probability density. If a variable $X_1$ is picked using a probability density $p_{X_1}(x)$ and is used to encode a message assuming an incorrect density $p_{X_2}(x)$, the amount of bits required is given by $H(X_1) + H(p_{X_1}||p_{X_2})$, where $H(X_1)$ is the classical Shannon entropy, instead of $H(X1)$ needed if the information about the distribution was correct. Thus the decrease of relative entropy in this case leads to a decrease in the uncertainty of the coding and increase in information about the process.
\end{thebibliography}
\end{document}